**On circular Bragg regimes in ellipsometry spectra of ambichiral sculptured thin films**

**Ferydon Babaei**

Department of Physics, University of Qom, Qom, Iran
Email: fbabaei@qom.ac.ir

**Abstract**

The generalized ellipsometry formalism is used for a right handed ambichiral sculptured thin film.The amplitude ratios and phase differences of this structure are extracted from the amplitude transmission ratios. The results showed that the circular Bragg regimes appear with abruptly changes in ellipsometry spectrum as wavelength, where the selective circular transmission is maximized.

## I. Introduction

The sculptured thin films (STFs) are three dimensional anisotropic nano-structures that can be produced by combination of oblique angle deposition and rotation of the substrate [1-5]. Substrate rotation causes that we can sculpture columns to arbitrary shapes and control thin film morphology. In chiral sculptured thin films (CSTFs), the substrate continuously rotates about its surface normal. The pitch (the structural period) of CSTF is determined by the rotation speed, the relative to the deposition rate, and the chiral handedness is set by the direction of substrate rotation [6]. A ambichiral sculptured thin film (ACSTF) is formed by discreet abrupt angular rotations($\varphi$ s in degrees) each time the film thickness has increased by $\frac{2\Omega}{n}$, where $2\Omega$ is the pitch of the helix and $n = \frac{360}{\varphi}$ is the number of sides of ACSTF [6, 7].

One of the interesting features of CSTFs is the occurrence of the circular Bragg phenomenon (or circular Bragg regime) in these films [8, 9]. By utilizing of the circular Bragg regime, use of CSTFs as circular polarization filters has been theoretically examined and then experimentally realized [10-13]. In the Bragg regime, CSTF filters reflect the circular polarization state of incident light matches the structural handedness of the material while opposite circular polarization state can freely propagate [14]. Therefore, in fabrication of CSTFs, substrate rotation as continuously or discretely and angular rotation will affect the circular Bragg regime.

Recently, generalized ellipsometry has been applied extensively for the study of optically anisotropic materials including organic thin films [15], STFs [16-18]. In anisotropic materials, the optical constants vary according to the propagation direction of light, and ellipsometry data analysis using conventional Fresnel equations becomes rather difficult [15]. In this study, in order to tracing the circular Bragg regimes in ellipsometry spectra, the amplitude transmission ratios and phase differences of an ACSTF have been calculated using the transfer matrix in conjunction with the generalized ellipsometry and compared with the those that obtained of a CSTF as a reference structure. The formalism for calculation these quantities outlined in section II. Numerical results are presented and discussed in Section III.

## II. Formulation

Consider a region $0 \leq z \leq d$ be occupied by an ACSTF (Fig.1). While the regions $z \leq 0$ and $z \geq d$ are vacuous. Let the structure be excited by a plane wave propagating at an angle $\theta_{inc}$ to the z- axis and at an angle $\psi_{inc}$ to the x- axis in the xy- plane. The phasors of incident, reflected and transmitted electric fields are given as [19]:

$$\begin{cases} \underline{E}_{inc}(\underline{r}) = [(\dfrac{i\underline{S}-\underline{P}_+}{\sqrt{2}})a_L - (\dfrac{i\underline{S}+\underline{P}_+}{\sqrt{2}})a_R]e^{i\underline{k}_0 \cdot \underline{r}}, & z \leq 0 \\ \underline{E}_{ref}(\underline{r}) = [-(\dfrac{i\underline{S}-\underline{P}_-}{\sqrt{2}})r_L + (\dfrac{i\underline{S}+\underline{P}_-}{\sqrt{2}})r_R]e^{-i\underline{k}_0 \cdot \underline{r}}, & z \leq 0 \\ \underline{E}_{tr}(\underline{r}) = [(\dfrac{i\underline{S}-\underline{P}_+}{\sqrt{2}})t_L - (\dfrac{i\underline{S}+\underline{P}_+}{\sqrt{2}})t_R]e^{i\underline{k}_0 \cdot (\underline{r}-d\underline{u}_z)}, & z \geq d \end{cases} \quad (1)$$

The magnetic field's phasor in any region is given as $\underline{H}(\underline{r}) = (i\omega\mu_0)^{-1}\underline{\nabla}\times\underline{E}(\underline{r})$, where $(a_L, a_R), (r_L, r_R)$ and $(t_L, t_R)$ are the amplitudes of incident plane wave, and reflected and transmitted waves with left- or right-handed polarizations. We also have:

$$\begin{cases} \underline{r} = x\underline{u}_x + y\underline{u}_y + z\underline{u}_z \\ \underline{k}_0 = k_0(\sin\theta_{inc}\cos\psi_{inc}\underline{u}_x + \sin\theta_{inc}\sin\psi_{inc}\underline{u}_y + \cos\theta_{inc}\underline{u}_z) \end{cases} \quad (2)$$

The unit vectors for linear polarization normal and parallel to the incident plane, $\underline{S}$ and $\underline{P}$, respectively are defined as:

$$\begin{cases} \underline{S} = -\sin\psi_{inc}\underline{u}_x + \cos\psi_{inc}\underline{u}_y \\ \underline{P}_{\pm} = \mp(\cos\theta_{inc}\cos\psi_{inc}\underline{u}_x + \cos\theta_{inc}\sin\psi_{inc}\underline{u}_y) + \sin\theta_{inc}\underline{u}_z \end{cases} \quad (3)$$

The reflectance and transmittance amplitudes can be obtained, using the continuity of the tangential components of electrical and magnetic fields at two interfaces, $z=0$ and $z=d$, and solving the algebraic matrix equation [19]:

$$\begin{bmatrix} i(t_L - t_R) \\ -(t_L + t_R) \\ 0 \\ 0 \end{bmatrix} = [\underline{\underline{K}}(\theta_{inc},\psi_{inc})]^{-1} \cdot [\underline{\underline{B}}(d,\Omega)] \cdot [\underline{\underline{M}}'(d,\Omega,\kappa,\psi_{inc})] \cdot [\underline{\underline{K}}(\theta_{inc},\psi_{inc})] \cdot \begin{bmatrix} i(a_L - a_R) \\ -(a_L + a_R) \\ -i(r_L - r_R) \\ (r_L + r_R) \end{bmatrix} \quad (4)$$

The different terms and parameters of this equation are given in detail (see equations (2-25), (2-26) in reference [19]). We now consider an ACSTF with thickness of $d = Nt_{sp}$, where $t_{sp} = \dfrac{\Omega\varphi}{180}$ is thickness of a side of ambichiral and N is the number of arms. The transfer matrix a columnar thin film with thickness of d is $e^{i[\underline{\underline{P}}]d}$ [20]. Therefore, the transfer matrix an ACSTF is:

$$[\underline{\underline{M}}]_{PCSTF} = [\underline{\underline{M}}]_N [\underline{\underline{M}}]_{N-1} \ldots [\underline{\underline{M}}]_3 [\underline{\underline{M}}]_2 [\underline{\underline{M}}]_1 \tag{5}$$

$$[\underline{\underline{M}}]_{i+1} = [\underline{\underline{B(\varphi)}}][\underline{\underline{M}}]_i [\underline{\underline{B(\varphi)}}]^{-1}, i = 1,2,\ldots,N-1 \tag{6}$$

, while $[\underline{\underline{B(\varphi)}}] = \begin{pmatrix} \cos\varphi & -h\sin\varphi & 0 & 0 \\ h\sin\varphi & \cos\varphi & 0 & 0 \\ 0 & 0 & \cos\varphi & -h\sin\varphi \\ 0 & 0 & h\sin\varphi & \cos\varphi \end{pmatrix}$ [21], $[\underline{\underline{M}}]_1 = e^{i[\underline{\underline{P}}]t_{sp}}$ and the

integers $h = \pm 1$ are for the structurally right- and left-handed ACSTF, respectively. We can rewrite Eq.5 using Eq.6 as:

$$[\underline{\underline{M}}]_{ACSTF} = [\underline{\underline{B(\varphi)}}]^{N-1} [\underline{\underline{M}}]_1 ([\underline{\underline{B(\varphi)}}]^{-1} [\underline{\underline{M}}]_1)^{N-1} \tag{7}$$

Finally, using the transfer matrix ACSTF in the Eq.7 (replace $[\underline{\underline{B}}(d,\Omega)] \cdot [\underline{\underline{M}}'(d,\Omega,\kappa,\psi_{inc})]$ by $[\underline{\underline{M}}]_{ACSTF}$ in Eq.4) one can obtain the reflection and transmission coefficients as $r_{i,j} = \frac{r_i}{a_j}$, $t_{i,j} = \frac{t_i}{a_j}$ ; $i, j = L, R, S, P$, where first and second indexes are devoted to reflected or transmitted and incident polarized plane wave, respectively. Selective circular transmission is $T_{LL} - T_{RR}$, so that circular reflection and transmission can be calculated as $R_{i,j} = |r_{i,j}|^2$, $T_{i,j} = |t_{i,j}|^2$ ; $i, j = L, R$.

In Spectroscopic ellipsometry of anisotropic thin films, the amplitude transmission ratios are normalized by $t_{ss}$ that can be obtained as [22, 23]:

$$\begin{cases} \tau_{pp} = \dfrac{t_{pp}}{t_{ss}} = \tan(\psi_{pp}) e^{i\Delta_{pp}} \\ \tau_{ps} = \dfrac{t_{ps}}{t_{ss}} = \tan(\psi_{ps}) e^{i\Delta_{ps}} \\ \tau_{sp} = \dfrac{t_{sp}}{t_{ss}} = \tan(\psi_{sp}) e^{i\Delta_{sp}} \end{cases} \tag{8}$$

, where $\psi$ and $\Delta$ include to amplitude ratio and phase difference of transmitted polarized plane wave.

### III. Numerical results and discussion

Consider a right-handed sculptured thin film (CSTF or ACSTF) in its bulk state has occupied the free space (Fig.1). The relative permittivity scalars $\varepsilon_{a,b,c}$ in this sculptured thin film were obtained using the Bruggeman homogenization formalism [24]. In this formalism, each column in the STF structure is considered as a string of identical long ellipsoids [20]. In all our calculations, the axial excitation ($\theta_{inc} = \psi_{inc} = 0^0$) of polarized plane wave is considered, columnar form factors $\gamma_\tau^s = \gamma_\tau^v = 20$, $\gamma_b^s = \gamma_b^v = 1.11$ (s and v, respectively indicate to inclusion and vacuum phase) and structural parameters of STF $\chi = 42°$ (rise angle), $2\Omega = 325 nm$, $f_v = 0.421$ (void fraction), $d = 30\,\Omega$ were fixed [6,25]. We have used the experimental data of the dielectric refractive index bulk titania ($TiO_2$) [26]. Also, we have included the dispersion and dissipation of dielectric function [27].

In order to find the circular Bragg regimes in ellipsometry spectrum of ACSTF, we have considered CSTF as a reference structure. The amplitude ratios and phase differences of CSTF with the inset plots of selective circular transmission are depicted in Fig.2. Circular Bragg regime occurs at 612nm with 82% maximum selective transmission with abruptly changes in ellipsometry spectrum as wavelength.

The amplitude ratios and phase differences for an ACSTF have depicted in Fig.3 at different abrupt angular rotations. We are increased angular rotation from 5° to 120° with a step 5° and ellipsometry spectra few of them are plotted in Fig.3. Until $\varphi = 70^0$, ellipsometry spectra of ACSTF similar results of those of CSTF as a reference structure are obtained. A CSTF contains a single circular Bragg regime centered at the Bragg wavelength $\lambda^{Br} \approx 2 n_{avg} \Omega$, where $n_{avg}$ is the average refractive index [6]. However, an ACSTF contains two circular Bragg regimes due to a stack of biaxial plates, a primary circular Bragg regime at $\lambda^{Br}$ and an inverted circular Bragg regime centered at $\lambda^{inv} \approx \frac{\varphi}{180^0 - \varphi} \lambda^{Br}$, where $\varphi$ is the angular rotation of the substrate between each arm of the ACSTF (in degrees)[6]. Our calculations showed that the intensity of secondary circular Bragg regime is negligible until $\varphi = 70^0$ and occurs at shorter wavelengths below 300nm. At $\varphi = 70^0$ inverted circular Bragg regime appears about 435nm with 62% maximum selective transmittance and with increasing angular rotation shifts to longer wavelengths. But for a tetragonal CSTF two circular Bragg regimes coincide at $\lambda^{inv} \approx \lambda^{Br}$ and obtained a residual selective transmission near to zero. This difference between our theoretical work and Popta *et al.'s* (2005) experimental work can be related to the structural difference between idealized theoretical model for ACTFs and that obtained in experimental work. In the latter, as Popta *et al.* (2005) have pointed out the experimental films exhibit a large amount of scattering due to the highly complex and non-ideal structure that the individual chiral elements exhibit. The results achieved in this work are consistent with the experimental data (Popta *et al.* (2005)). Therefore, two circular Bragg regimes can be adduced from spectral signatures as abruptly changes in ellipsometry spectrum in these thin films.

## IV. Conclusions

In this work, we theoretically analyzed the amplitude ratios and phase differences of ACSTFs using generalized ellipsometry formalism. In comparison to CSTFs, the results showed that in the lower angular rotations ($\varphi < 70^0$) do not exist difference between ellipsometry spectra of ACSTFs and CSTFs. However, the difference appears at higher angular rotations as two circular Bragg regimes that there exists good compatible to selective circular transmission. The results of this work may be applied to the ellipsometry of chiral sculptured thin films.

**Acknowledgements**

We wish to acknowledge support from the University of Qom.

**Figure captions**

Fig.1. Schematic of a single column of (a) a chiral STF (b) a trigonal STF and (c) the top view of a, b.

Fig2. Calculated the amplitude transmission ratios ($\psi_{pp}$, $\psi_{ps}$ and $\psi_{sp}$) and phase differences ($\Delta_{pp}$, $\Delta_{ps}$ and $\Delta_{sp}$) spectra of a right handed-CSTF. The inset plots the selective transmission of circularly polarized light.

Fig.3. Same as Fig. 2, except that spectra is calculated for a right handed- ACSTF at different angular rotations.

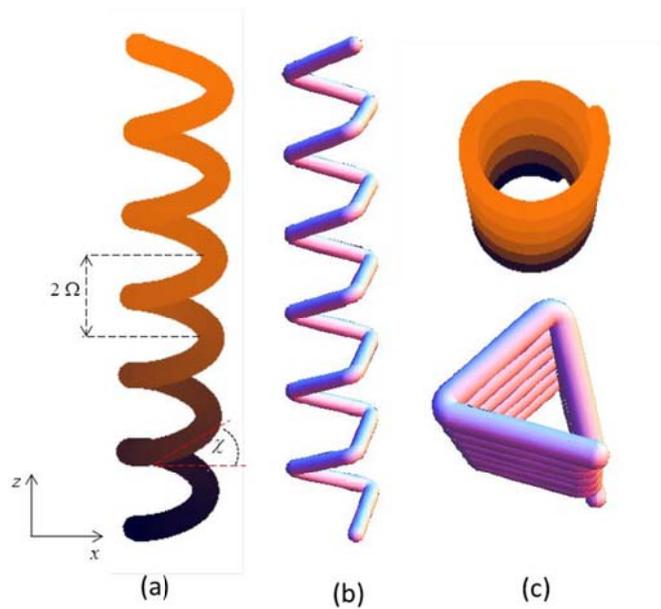

**Fig. 1; F. Babaei**

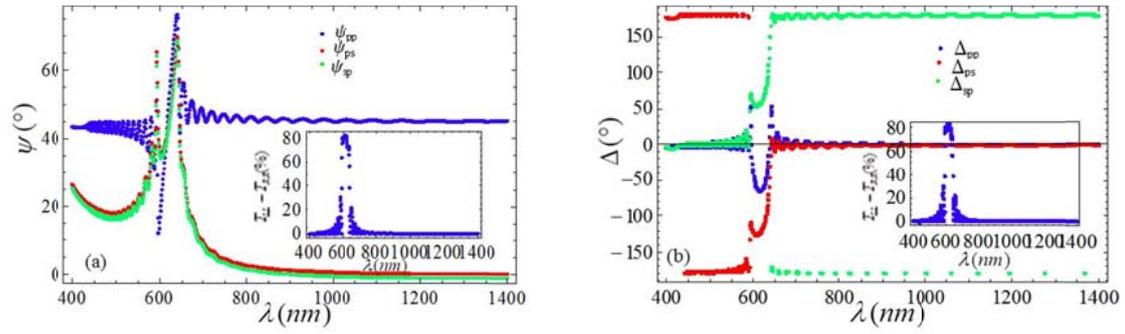

**Fig. 2; F. Babaei**

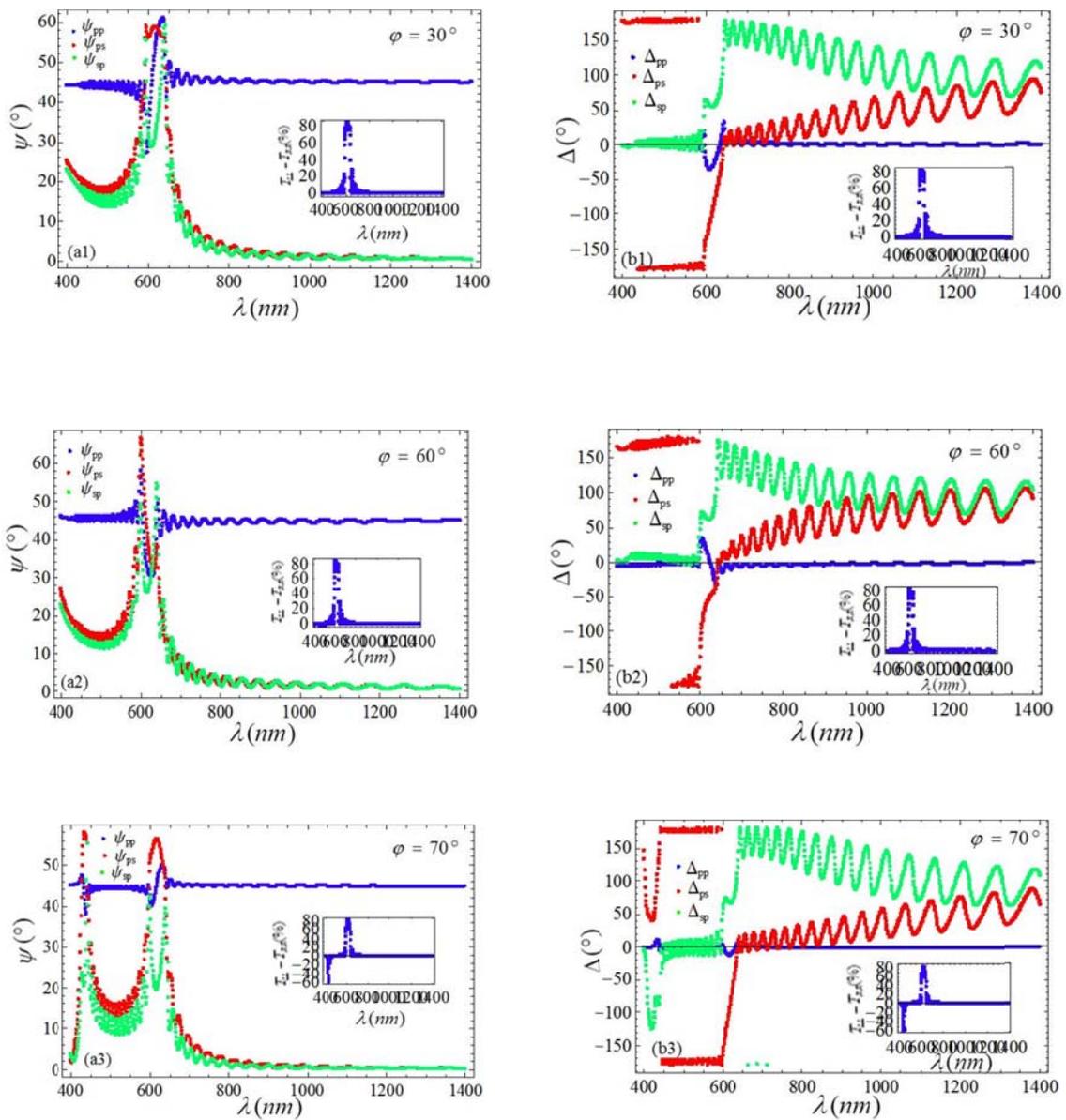

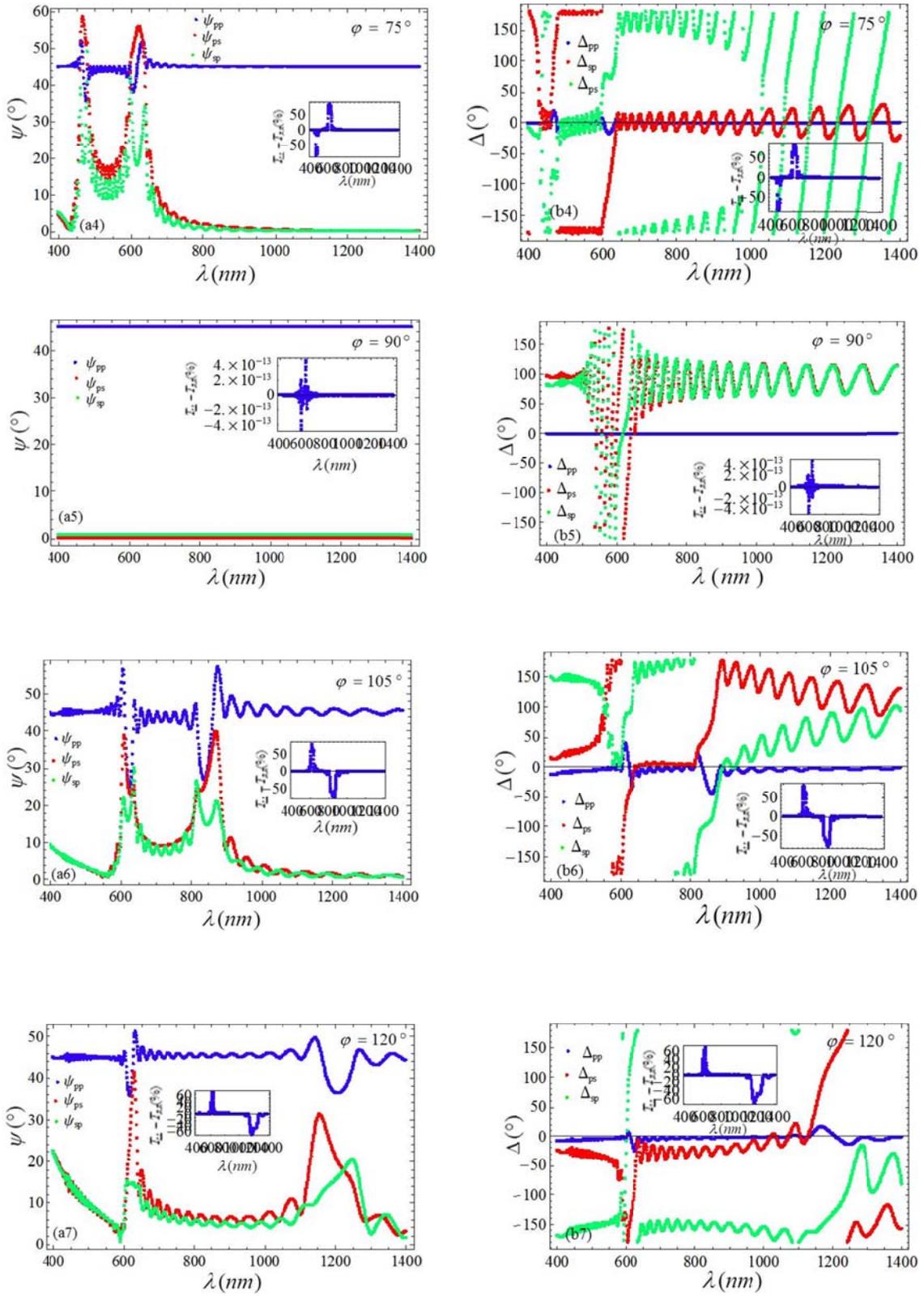

**Fig. 3; F. Babaei**